# Anisotropy of thermal transport in phosphorene: A comparative first-principles study using different exchange-correlation functional

Fa Zhang[1,2], Xiong Zheng[2], Huimin Wang[3], Liang Ding[1*], Guangzhao Qin[2*]

[1] *State Key Laboratory of Robotics and System, Harbin Institute of Technology, Harbin 150001, People's Republic of China.*

[2] *State Key Laboratory of Advanced Design and Manufacturing for Vehicle Body, College of Mechanical and Vehicle Engineering, Hunan University, 410082 Changsha, P. R. China.*

[3] *Hunan Key Laboratory for Micro-Nano Energy Materials & Device and School of Physics and Optoelectronics, Xiangtan University, Xiangtan 411105, Hunan, China.*

*Email: liangding@hit.edu.cn; gzqin@hnu.edu.cn

***Abstract***: Phosphorene, as a new type of two-dimensional (2D) semiconductor material, possesses unique physical and chemical properties, and thus has attracted widespread attention in recent years. With its increasing applications in nano/optoelectronics and thermoelectrics, a comprehensive study on its thermal transport properties is necessary. It has been concluded from previous studies that there exist vast difference and uncertainty in the theoretically predicted thermal conductivity, which is generally attributed to the selection of XC functional. However, there is no comprehensive investigation on this issue. In this paper, based on first-principles calculations using 12 different exchange-correlation (XC) functionals, the phonon transport properties of phosphorene are systematically studied by solving the Boltzmann transport equation (BTE). The results show noticeable differences in the phonon transport properties of phosphorene under different XC functionals. For instance, the thermal conductivity of phosphorene along the zigzag direction ranges from 0.51 to 30.48 $Wm^{-1}K^{-1}$, and that along the armchair direction ranges from 0.15 to 5.24 $Wm^{-1}K^{-1}$. Moreover, the anisotropy of thermal conductivity is between 3.4 and 6.9, which is fundamentally originated from the special hinge-like structure. This study conducts an in-depth analysis of phonon transport to reveal the effect and mechanism of different XC functionals for predicting thermal transport properties, which would provide a reference for future research on phosphorene and other novel materials.

# 1. Introduction

Since the discovery of graphene in 2004, two-dimensional (2D) materials have come into view [1]. The low-dimensional physicists are making breakthroughs on the basis of the concept of graphene, which represents a new type of atomically thick materials [2]. During recent years, 2D black phosphorene has gotten more and more attention with the development of science and technologies [3]. Black phosphorus is a layered material in which individual atomic layers are stacked together by *van der Waals* interactions, just like graphite [4]. However, a single layer of black phosphorus (the phosphorene) is a semiconductor with direct bandgap, which is different from graphene with Dirac cone [5, 6]. Each phosphorus atom is covalently bonded to three adjacent phosphorus atoms, forming a folded honeycomb structure [5-7]. Furthermore, phosphorene has exceptionally high hole mobility (on the order of 10,000 $cm^2V^{-1}s^{-1}$) and unusual elastic properties, which make phosphorene a promising candidate for the source material of field-effect transistors [8]. Besides, phosphorene is considered as the potential thermoelectric material, which requires high electrical conductivity and low thermal conductivity [9]. It was found that the anisotropy of phosphene's electrical and thermal transport is orthogonal to each other [10,11]. Generally speaking, thermoelectric performance and efficiency can be characterized by Seebeck coefficient, $ZT = S^2\sigma T/\kappa$ [12]. By applying a temperature gradient along the direction of the phosphorene armchair, the maximum electrical conductivity (σ) and the minimum thermal conductivity($\kappa$) can be obtained at the same time, thereby obtaining the maximum ZT value. Thus, phosphorene is promising in the field of thermoelectrics [13-20]. Considering the potential values of phosphorene in the applications of nanoelectronics and thermoelectrics, the comprehensive studies on its thermal transport properties, especially the characteristics of anisotropy, are on demanding.

The thermal transport properties play a vital role in lots of applications as mentioned above.. Even though the thermal conductivity of phosphorene and bulk phosphorus has been experimentally characterized well, the thickness of the phosphorene samples for measuring thermal conductivity is too large to reach the ideal state due to the limitation of synthesis technology [21,22]. Thus, in addition to experimental measurements, researchers tend towards the theoretical prediction of the thermal conductivity of phosphorene and the exploration of phonon transport properties [23-26]. While the thermal conductivity of phosphorene is still ambiguity even if there have been a tremendous studies in literature. There are few possible reasons for this ambiguity. First, different empirical potential functions or force fields are fitted following different procedures and are used to describe the interatomic interaction in classical molecular dynamics (MD) simulations, which have a significant effect on the results. Second, lots of pseudopotentials and exchange-correlation (XC) functionals are available in the first-principles, which are constructed in different frame, such as the local density approximation (LDA) and the generalized gradient approximation (GGA). The different pseudopotentials and XC functionals have a significant effect on the evaluation of thermal conductivity. For instance, XC functional groups affect the optimization of the crystal structure and the lattice constant, which have a decisive influence on the lattice vibration and heat transfer performance [27]. In previous studies, researchers have found a vast difference in the thermal conductivity of single-layer phosphorene. Specifically, the thermal conductivity of phosphorene along the zigzag direction is between 15.33 and 152.7 $Wm^{-1}K^{-1}$, the thermal conductivity in the armchair direction is between 4.59 and 63.9 $Wm^{-1}K^{-1}$[28-29]. Previous studies attributed the large uncertainty of thermal conductivity of phosphorene to the selection of XC functions. Even though the selection of functional groups in the first-principles calculations is particularly essential for accurately predicting the thermal conductivity of phosphorene, there is no comprehensive investigation on this issue. It is well known that black phosphorene is entirely different from planar graphene and flexural silicene crystal structure [27, 30–32]. 2D phosphorene has a hinge-like structure and is folded along the armchair direction. The hinge-like structure leads to different atomic arrangement periods and different degrees of density in different directions, which results in the fantasic anisotropy in thermal transport properties of phosphene. Thus, the goal of this study is to systematically investigate the effects of different XC functions on the thermal conductivity of phosphorene.

In this paper, the thermal transport properties of phosphorene are quantitatively studied by solving the phonon Boltzmann transport equation (BTE) based on first-principle calculations using different exchange-correlation (XC) functions. The results show that the anisotropy in the thermal transport process of phosphene exhibits significant differences with different XC functions. Moreover, the thermal conductivity in the zigzag direction ranges from 0.51 to 30.48 $Wm^{-1}K^{-1}$ and the thermal conductivity in the armchair direction is between 0.15 - 5.24 $Wm^{-1}K^{-1}$. In-depth study on phonon behavior is conducted to analyze the mechanism underlying the anisotropy of thermal transport in phosphorene. The results achieved in this study provide valuable reference for the accurately quantitative calculation of phosphorene and future research of other novel materials, which is of considerable significance to nano-engineering applications.

**2. Computational details**

All the first-principles calculations are performed based on density functional theory (DFT) using the Vienna *ab initio* simulation software package (VASP) [33]. For the structure optimization and following DFT calculations of phosphorene, different poeudopotentials[34, 35]and XC functionals in the LDA and GGA are employed: LDA [36], Perdew-Burke-Ernzerhof (PBE) [37], revised PBE (rPBE) [38], PBE revised for solids (PBEsol) [39], Perdew-Wang 91 (PW91) [40], ultrasoft pseudopotential GGA (USPP_GGA), ultrasoft pseudopotential LDA (USPP_LDA) [41], the "opt" functionals (optB88, optB86b, optPBE) [42, 43], the vdW_DF2 and the vdW_DF22 [44].

Based on the full convergence test, the cutoff values of the kinetic energy of the wave function set to 700 and 450 eV for PAW and ultrasoft (USPP_GGA and USPP_LDA) pseudopotentials, respectively. A Monkhorst–Pack *k*-mesh of 15×11×1 is used to sample the Brillouin Zone (BZ), and the energy convergence threshold is set to $10^{-8}$ eV. In order to prevent the interaction caused by periodic boundary conditions, a vacuum layer of 20 Å is used along the *out-of-plane* direction. To completely optimize the shape and volume of the unit cell, all the atoms have been relaxed until the maximum Hellmann-Feynman force acting on each atom is not higher than $10^{-8}$ eV/Å.

Before calculating the second-order and third-order interatomic force constants, the unit cell is need to be expanded to predict the dynamics of the lattice accurately. By judging the convergence of phonon dispersion, the supercell size of phosphorene is chosen as 6×6×1. For the calculation of all IFCs in phosphorene, space group symmetry is used to reduce the calculation cost and the numerical noise of IFCs [45]. Thoroughly test the cutoff distance (r-cutoff) in the anharmonic IFC calculation, and the specific results can be referred to as the analysis and discussion. A detailed discussion on convergence will be given later based on Figure 2 and Supporting information. The use of Lagrangian multiplier method [46,47] strengthens the translation and rotation invariance of IFCs. Based on the density functional perturbation theory (DFPT), the dielectric constant (has been marked in Table 1) is obtained. $\kappa$ is obtained by using an iterative process to solve the linearized phonon BTE based on force constants, which is implemented in the ShengBTE package [27, 46].

## 3. Results and discussions

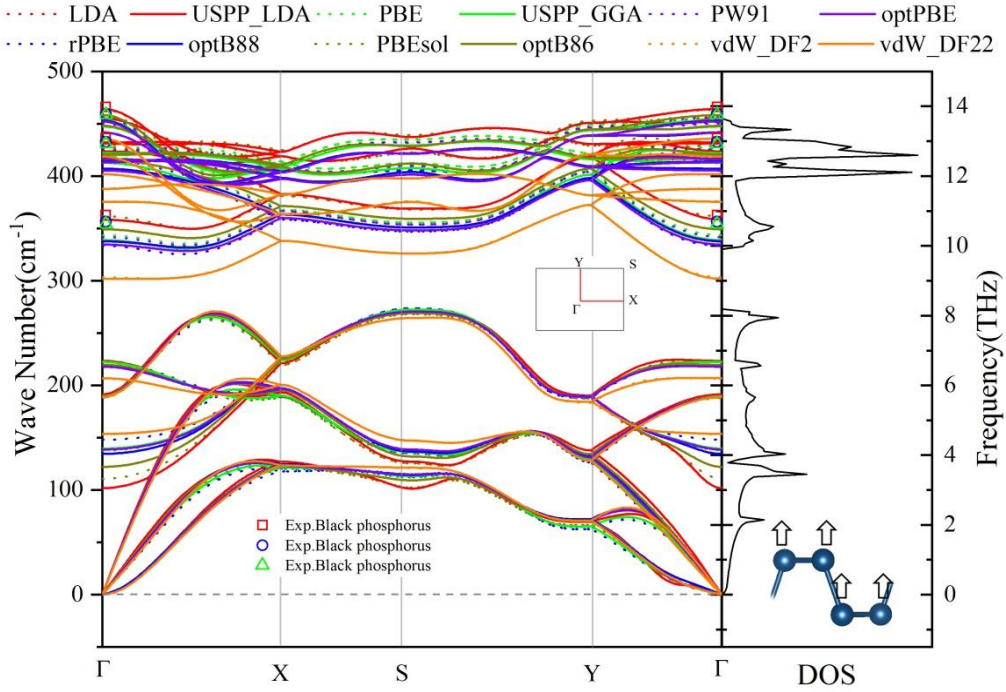

**Fig. 1.** Comparison of the phonon dispersions of 2D black phosphorene calculated by different XC functional. The experimentally measured data of bulk black phosphorus[26, 27, 48] are also plotted. The drawn phonon density of states (DOS) are calculated using the optB86b functional. Inset: the IBZ and the side view of phosphorene where arrows mark the vibration mode of FA phonon branch.

Based on the optimized structure, the phonon dispersion map is calculated by phonopy as shown in Figure 1, where the results calculated using different XC functions are collected together for the comparison. There is no imaginary frequency in all the optimized structures, which means the structures are thermodynamically stable. Therefore, in the following discussions, we will compare and discuss the thermal transport properties of phosphorene calculated by 12 different kinds of XC functions. Because only the Raman scattering curve of bulk black phosphorus can be obtained from previous experiments, we only present the comparison of the optical phonon branches at Γ point in Fig. 1. It can be clearly seen that the FA phonon branch calculated using the LDA function based on the super-soft pseudopotential shows an apparent softening phenomenon. The rest of the XC functions are consistent for the acoustic phonon modes. However, the main difference occurs in the optical phonon modes, which can be compared based on the frequency of the $A_g^1$, $A_g^2$, and $B_{2g}$ phonon branches at the BZ center (Γ point). According to the experimental measurement, the frequency is between 359.51 and 368.21 cm$^{-1}$ for $A_g^1$, 439. 61 and 443. 03 cm$^{-1}$ for $A_g^2$, and 465.85 and 473.43 cm$^{-1}$ for $B_{2g}$. The corresponding points are marked in Fig. 1. All the calculations agree very well with the experiments except the optical mode of vdW_DF2/vdW_DF22, which is significantly lower than the experimental values.

According to phonon spectrum analysis from Fig. 1, the phonon modes below the band gap play a decisive role in the thermal transport. Moreover, these modes dominate the anisotropy of heat transfer. Readers can see the detailed discussion presented in Fig. 3.

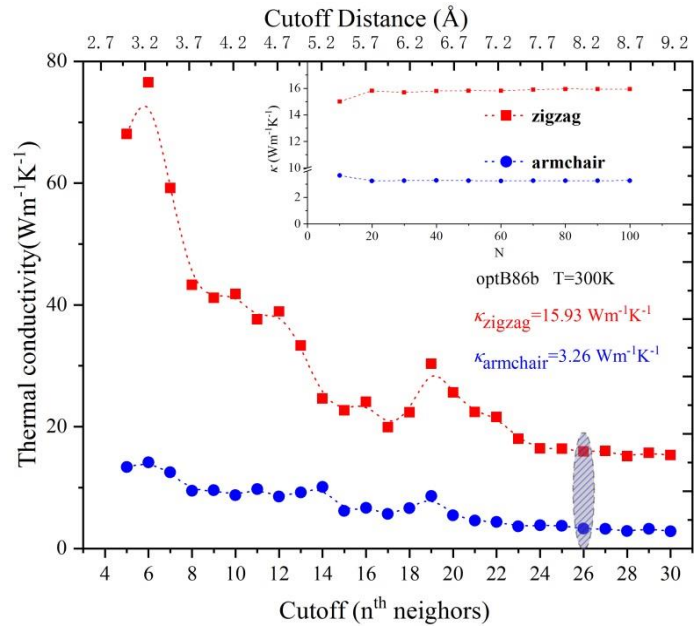

**Fig. 2.** The convergence behavior of thermal conductivity ($\kappa$) with respect to the cutoff and $Q$-grid ($N \times N \times 1$). The presented results are calculated using the representative optB86b. The cutoff for the converged thermal conductivities is highlighted onsite.

In the past few years, first-principles calculations combining with BTE methods are recognized as good methods to predict the heat transport properties of materials. While there are still some factors that can affect the final calculation results. One of the important factors is cutoff radius of the third-order force constant and the convergence of the $Q$-mesh division, in order to obtain accurate thermal conductivity prediction results, they should be carefully judged. As shown in Figure 2 (additional material 1), under the condition that the phonon-phonon scattering caused by anharmonicity is fully respected, we have done the close truncation and Q-grid convergence tests on all 12 XC functionals.

The results show that for all XC functions, the thermal conductivity of black phosphorene has a good convergence. According to the judgment of convergence, the cutoff radius for different XC functions are as follows: USPP_LDA (7.3Å), optB86/LDA (7.7 Å), USPP_GGA/optPBE (8.0 Å), and other XC functionals (8.2 Å). Besides, for the convenience of comparison, all XC functional Q grids are selected as 100×100×1.

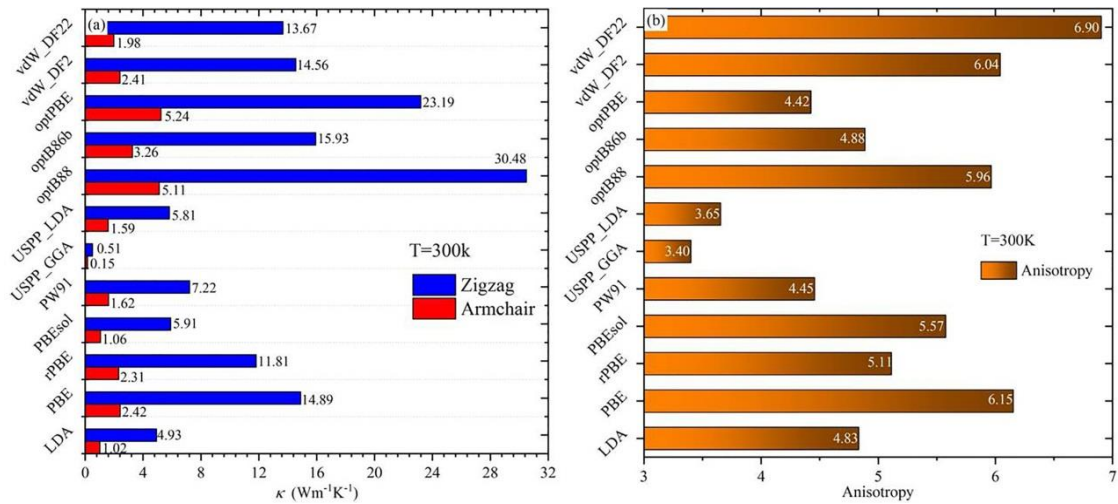

**Fig. 3.** (a) Thermal conductivity of black phosphorene at 300K along the zigzag and armchair directions calculated using different XC functional. (b) Anisotropy ratio of thermal conductivity of black phosphorene at 300K, which is defined as $\kappa_{zigzag}/\kappa_{armchair}$.

It can be seen in Fig. 3(a) and (b) that there are differences in the predicted values of thermal conductivity using different functions. However, the common conclusions on the properties of heat transport inside the phosphorene can be seen. The thermal conductivity of single-layer phosphene is lower in both the zigzag direction and the armchair direction, both of which are two orders of magnitude lower than the thermal conductivity of graphene (3000-5000Wm$^{-1}$K$^{-1}$) [33, 47, 48 ]. This is because the FA phonon mode contributes less to the two thermal conductivities in black phosphorene. The scattering rate of the FA phonon branch in the phosphorene is very high, resulting in a small contribution of FA to the thermal conductivity.

Fig. 3(a) shows that the thermal conductivity calculated by the "opt" function (optB88, optB86b, optPBE) and the "vdW" function (vdW_DF2, vdW_DF22) is higher. This is because these XC functions include van der Waals (vdW) interactions. Due to the special hinge structure of the monolayer phosphene, many adjacent P atoms do not form bonds, but there are still some interactions. These effects can be captured by the XC functions including vdW interactions, the vdW interaction has a significant impact on the properties of bulk BP and phosphorene. The vdW interaction will disrupt the formation of resonance bonds by changing the lattice constant, especially the distance between the non-covalent bond P atoms, leading to long-range interactions. Therefore, the thermal conductivity calculated from "opt" function and "vdW" function is higher under the same cutoff radius. From Fig. 3(b), it can be seen that the transport of phosphene has a strong anisotropy. Previous research results show that the thermal conductivity of phosphene in the zigzag direction is greater than that in the armchair direction, and the ratio ranges from 2.2 to 5.5 [27]. The figure shows vdW_DF2, vdW_DF22, optB88 (considering the vdW interaction), the anisotropy ratio of PBE (without considering the vdW interaction) is higher than the value interval previously measured. It can be seen that, although the prediction of thermal conductivity will be higher when considering the interaction of vdW, it has little effect on anisotropy.

The anisotropy of thermal conductivity is attributed to the anisotropy of phonon group velocity determined by phonon dispersion [23, 50]. As shown in Fig. 1, the thermal conductivity in the armchair direction of black phosphorene is mainly contributed by the LA phonon mode. For the thermal conductivity in the zigzag direction, TA/FA participates in addition to the contribution of the LA phonon mode, the contribution of direction FA is more significant than that of armchair direction. The reason is that the peculiar hinge structure of phosphorene causes the "uneven" in the armchair direction, which leads to the difference of the phonon mode in the two directions, and finally makes the thermal conductivity anisotropy.

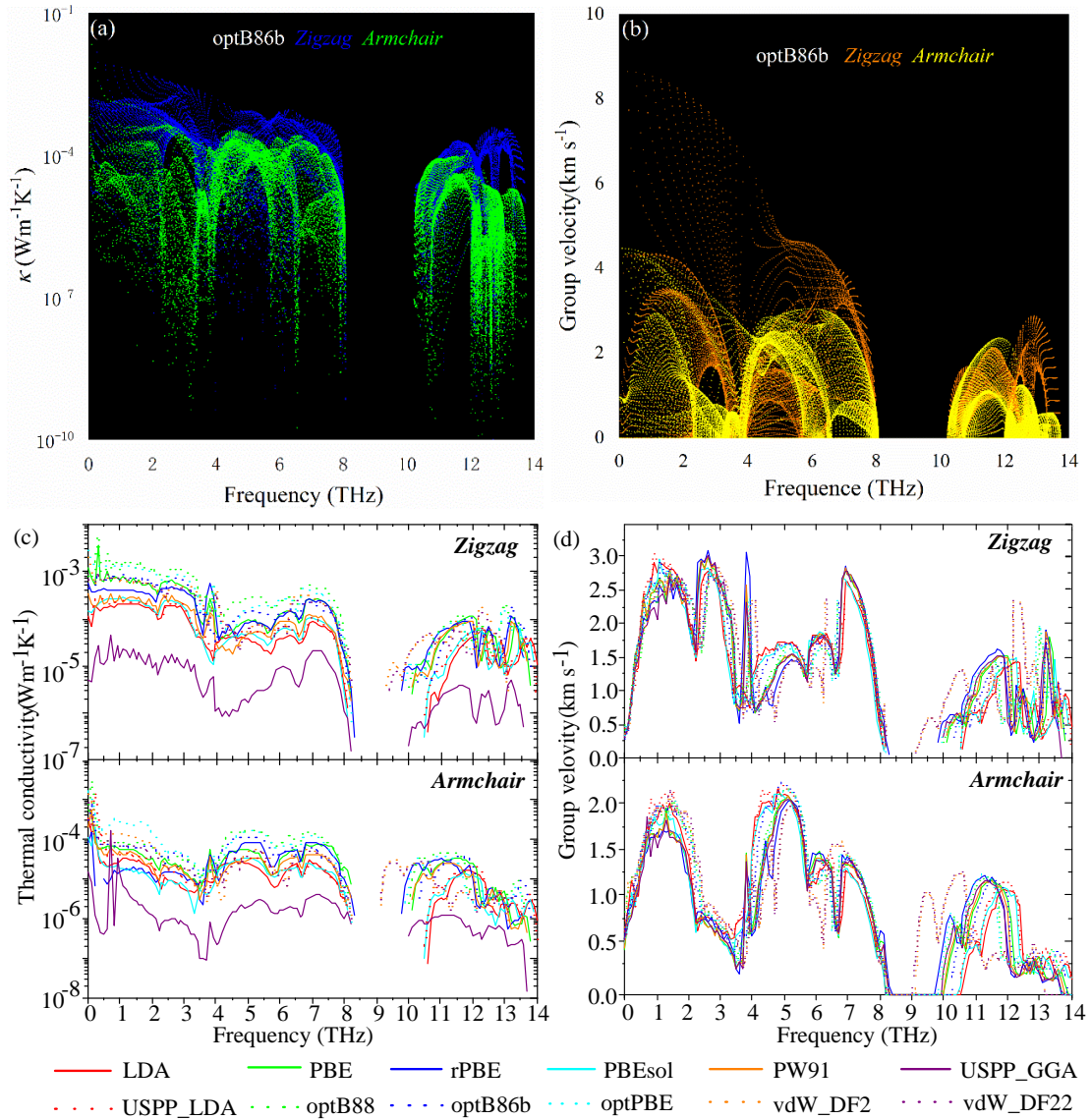

**Fig. 4.** The modal contribution to (a) thermal conductivity and (b) group velocity as a function of phonon frequency at 300 K for the representative functional of optB86b. (c,d) The variation of (c) thermal conductivity and (d) group velocity with respect to phonon frequency at 300 K for 12 different

functionals. The results show the weighted average of the thermal conductivity and group velocity of all phonon modes in a fixed frequency range.

To observe the anisotropy in the heat transport process of phosphorene more intuitively, the modal contribution is further examined as shown in the Fig. 4(a-b). The XC function used here is the representative optB86b. Fig. 4 (a) clearly shows that the acoustic phonon branch is the main contributor to the thermal conductivity of phosphorene. Fig. 4(b) shows that the phonon group velocity in the zigzag direction in the acoustic phonon branch is significantly higher than in the armchair direction. In the additional information, we have given the relationship among the phonon frequency of the other 11 functions, the thermal conductivity, and group velocity.

Fig. 4(c-d) shows the influence of different XC functions on thermal conductivity and group velocity. In Fig. 4(c), the thermal conductivity of USPP_GGA in both zigzag and armchair directions is abnormally low. This is consistent with the result reflected in Fig. 3(a). It is worth noting that although Fig. 1 shows that the phonon spectrum using USPP_GGA has no imaginary frequency indicating the stability of thermodynamics, this does not mean that USPP_GGA can accurately predict the thermal conductivity of the materials. The reason may be that the features of USPP and GGA are not compatible. Therefore, we suggest that the use of USPP_GGA functional groups should be avoided when calculating the thermal transport properties of black phosphorene. From the group velocities shown in Fig. 4(d), it can be observed that the group velocities of the acoustic phonon branches of vdW_DF2 and vdW_DF22 in the armchair direction are smaller than all other functions, and the zigzag direction is more significant than other functions. This makes the thermal conductivity calculated from vdW_DF2 and vdW_DF22 in the armchair direction is small, and the thermal conductivity in the zigzag direction is relatively large. Therefore, the thermal conductivity anisotropy ratio of these two XC functions is higher than that of all other functions.

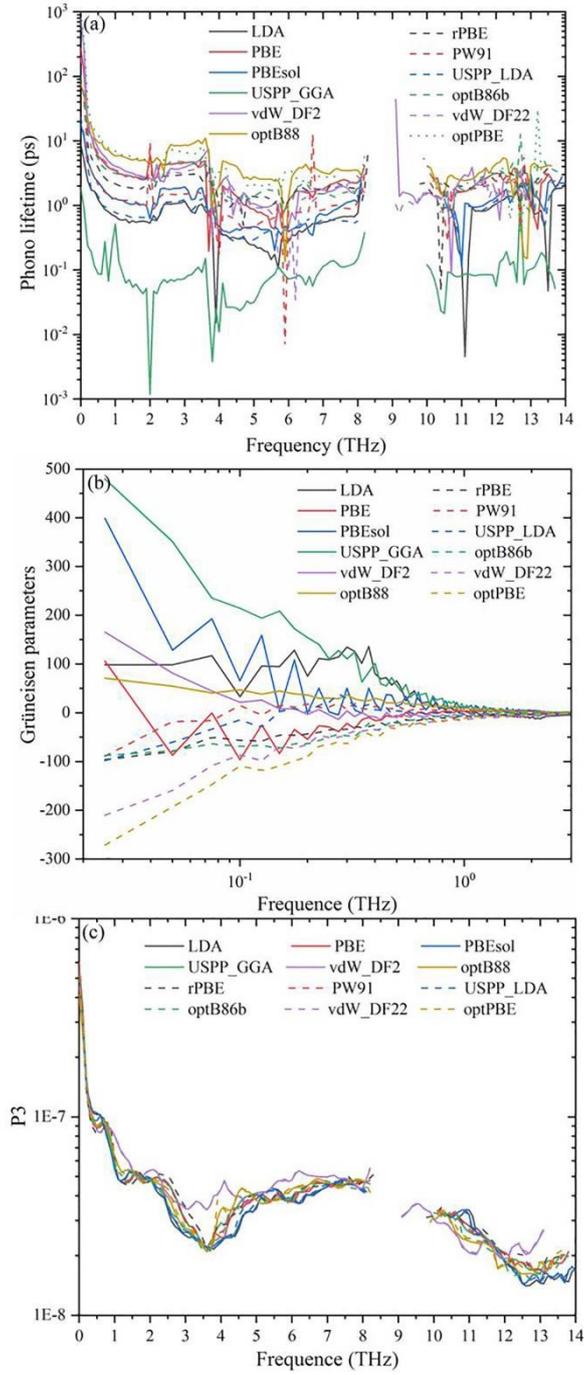

**Fig. 5.** The frequency resolved phonon lifetime, Grüneisen parameter, and scattering phase space. (a) Comparison of the phonon lifetime of phosphorene calculated using different XC functionals at 300 K. (b) The Grüneisen parameters in low frequency area of 0-2THz where the difference mainly emerges. （c）The variation of scattering phase space with respect to phonon frequency at 300K for the 12 different functional.

The understanding of thermal conductivity can be deepened through the analysis of phonon lifetime. The phonon lifetime predicted by optPBE and optB88 is above the rest of the XC functions. This is the reason for the high thermal conductivity of optPBE and optB88. As we all know, the thermal conductivity of micro-nano materials is affected by three factors, which are specific heat capacity ($Cv$), group velocity ($v$), phonon lifetime ($t$) [51]. Among them, the specific heat capacity and group velocity are wholly affected by the harmonic characteristics, while the phonon lifetime is affected by the scattering phase space and scattering intensity. In Table 1 and Fig. 5(c), we can see that the scattering phase space is similar. Now it is imperative to consider the scattering intensity, that is, the phonon anharmonicity quantified by the Grüneisen parameter.

Fig. 5(b) shows the comparison of Grüneisen parameters after using different XC functions. As can be seen in Fig. 5(b), the Grüneisen parameters from optB88 is relatively small at 300K, which results that the thermal conductivity of black phosphorene from optB88 is significantly higher than which from other functions. LDA, PBEsol, USPP_GGA have larger parameter values, which causes a smaller thermal conductivity using these XC functions. One thing should be noted is that the thermal conductivity is derived from the coupling of three factors, that is to say, the smaller the Grüneisen parameter, the longer the phonon lifetime obtained using different functions in phosphorene, but not necessarily resulting in higher thermal conductivity. Table 1 shows that the total Grüneisen parameter of optB86b is the smallest, but the thermal conductivity obtained at 300K is not the largest.

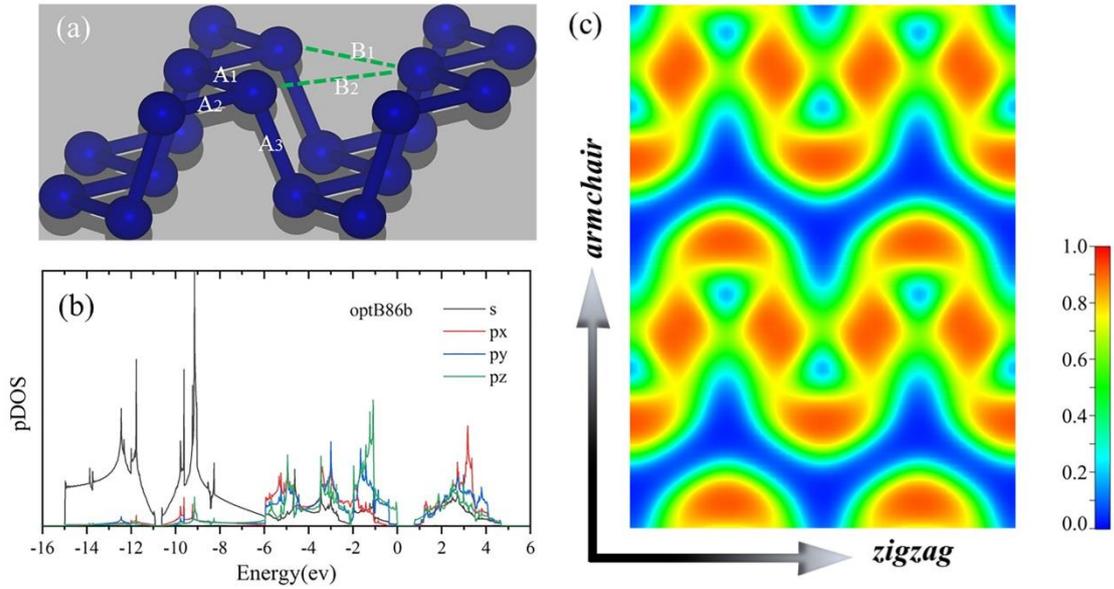

**Fig. 6.** (a) The perspective view of the geometric structure diagram of phosphorene. $A_1$-$A_3$ mark bonds, and the dashed lines ($B_1$, $B_2$) indicate the resonant interactions between the non- covalently bonding P atoms. (b) The orbital ($s$, $p_x$, $p_y$, $p_z$) projected density of states (pDOS) of electrons. (c) Top view of electron localization function (ELF).

Phosphorene has a hinge-like structure, as shown in Fig. 6a, and its wrinkled structure can actually be seen as a deformation of a two-dimensional planar structure [27]. Each P atom has five bonds formed. However, since the *sp*-hybridization in phosphene is weak [53], only three *p* electrons can be used for bonding, which leads to unsaturated covalent bonding in phosphene. Therefore, the true bond state in phosphene is the hybridization or resonance between the different electronic configurations where the electrons occupy the *p* orbital. In fact, compared with the resonance bonding of the conventional rock salt structure, the resonance bonding in the phosphorene exists in a weakened form [54]. This is due to the structural deformation of the perfect rock salt structure of the phosphorene hinge-like structure. However, the weakened resonance bond still has a non-negligible effect on the phonon transport properties of phosphorene. The stronger the resonance bond is, the lower the thermal conductivity is [55]. As shown in Fig. 6(a), due to the deformation, the first nearest neighbor distance in phosphorene is A1 and A2, and the second nearest neighbor distance is A3. As shown in Table 1, the optimized structure of USPP_LDA has the smallest first and second nearest neighbor distance (A1=A2=2.196Å, A3=2.224Å). Therefore, the resonant bonding is further enhanced in USPP_LDA. This structural deformation further enhances long-distance interaction, which makes the thermal conductivity calculated by the USPP_LDA XC functional becoming lower.

The armchair and zigzag directions show significantly different bonding characteristics. This is another explanation for the anisotropy of thermal conductivity. The electron ($s$, $px$, $py$, $pz$) orbital projection density of states (pdos) is shown in Fig. 6(b). From the calculated results, it can be seen that the bonding state near VBM is mainly controlled by the $pz$ orbitals hybridized with the $px$ and $py$ orbitals. The $s$ orbital is mainly restricted to around 9eV below the VBM, showing a weak hybridization relationship with the $px$, $py$ and $pz$ orbitals.

Fig. 6(c) shows the in-plane electron density of phosphorene. Anisotropic behavior can also be understood from the basic point of view of atomic bonds. In order to depict the probability of occurrence of an electron pair, we plot the ELF of phosphorene in Fig. 7(c) [10, 52]. ELF shows the position and size of the bond and lone pair electrons. ELF ranges from 0 to 1, where 0 means no electrons, and 1 means the position with the greatest degree of electronic localization. The ELF of the phosphorene along the zigzag direction is greater than 0.5, which means that the electrons are localized. For the armchair direction, the ELF is less than 0.5, which means that the electrons are delocalized. The ELFs along the two directions show great differences for the phosphorene, which is the physical reason for the anisotropy of a series of elements with a hinge-like structure.

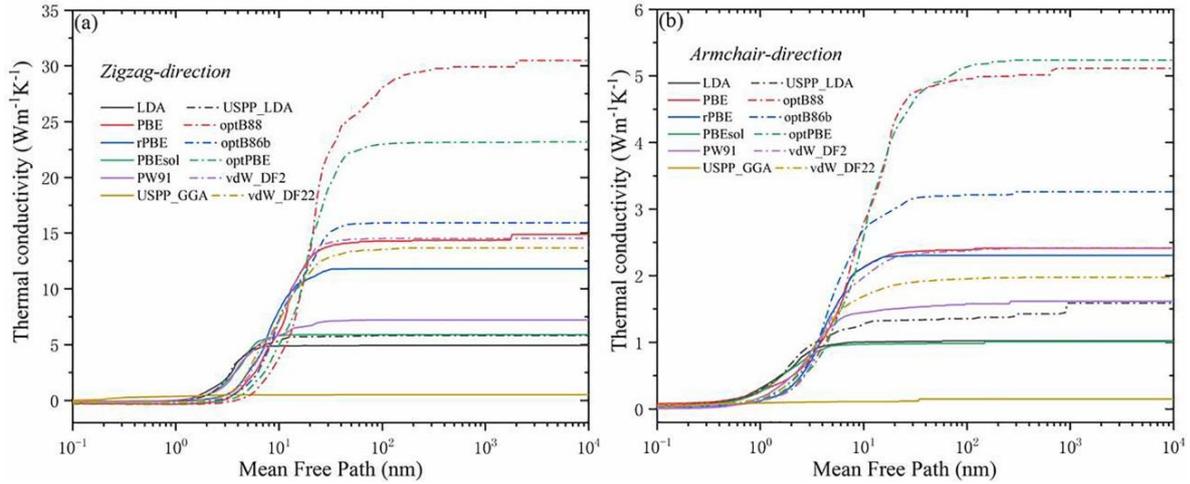

**Fig. 7.** The cumulative thermal conductivity of phosphorene along (a) zigzag and (b) armchair directions with respect to the phonon MFP for different XC functionals at 300 K.

The size of the nanostructure is smaller than the feature size, which can effectively control the thermal conductivity. So, the measurement of the MFP is an essential consideration in the thermal design of the nanostructure. Fig. 7 plots the cumulative thermal conductivity ($k$) of the phonon free path in the zigzag direction and armchair direction at 300 k. Comparing Fig. 7 (a) and Fig. 7 (b), when the size is $1\times10^4$ nm, all the phosphorene $\kappa$ calculated by the XC function are all converged. On the whole, due to the anisotropy of black phosphorene, the MFP in the zigzag direction is more substantial, which causes the zigzag direction $\kappa$ to rise more drastically with the MFP. In order to understand ballistics and phonon transport better, Table 1 lists the MFP when $k$ is 50%. For most XC functions, the MFP in the zigzag direction is between 2 and 10 nm, and the MFP in the armchair direction is between 2 and 4 nm. After comparison, the MFP calculated by optPBE, optB88 and PBE is higher than the average value. The MFP obtained by LDA function calculation is much smaller than the average value.

## 4. Conclusions

In summary, 12 different XC functionals combined with super soft potentials or PAW pseudopotentials were used to study the thermal transport properties of black phosphorene based on first-principles calculations. The thermal transport properties of phosphorene show distinct differences with different XC functionals. The results show that with different XC functional groups, the thermal conductivity of phosphorene in the zigzag direction is between 0.51 and 30.48 $Wm^{-1}K^{-1}$, and the thermal conductivity in the armchair direction is between 0.15 and 5.24 $Wm^{-1}K^{-1}$. Specifically, the XC function considering the interaction of vdW can achieve higher thermal conductivity. The difference lies in the lifetime of the low-frequency phonon mode near the center of BZ. Moreover, for different XC functionals, the ratio of anisotropy of thermal conductivity is between 3.40 and 6.90. Among them, vdW_DF2, vdW_DF22, PBE predict higher the anisotropy. It shows that vdW interaction is not the main factor directly affecting the anisotropy. Just like the thermal conductivity calculated by the USPP_LDA function, the strengthening of the resonance bond will cause the thermal conductivity to decrease. In this work, a comprehensive study of the thermal transport properties of black phosphorene with different XC functional groups is carried out, which provides a reference for future research on phosphorene and other new materials.

**Table. 1.** Calculated properties of black phosphorene at 300 K, including lattice constant, lattice thermal conductivity ($\kappa$), dielectric constants ($\varepsilon$), mean free path (MFP), Grüneisen parameter, specific heat capacity, three-phonon scattering phase space, and using different XC functionals, count the length of the A1-A3 bond, and the length of B1, B2 that did not form a bond.

| XC | Direction | Lattice constant (Å) | $\kappa$ (Wm$^{-1}$K$^{-1}$) | Dielectric constant | MFP (nm) | Grüneisen parameter | Heat capacity ($10^5$ Jm$^{-3}$K$^{-1}$) | Phase space ($10^{-3}$ a. u.) | $A_1=A_2$ | $A_3$ (Å) | $B_1=B_2$ |
|---|---|---|---|---|---|---|---|---|---|---|---|
| LDA | Zigzag | 3.267 | 4.93 | 4.373 | 2.943 | 1.281 | 18.094 | 3.299 | 2.199 | 2.225 | 3.324 |
| | Armchair | 4.367 | 1.02 | 8.864 | 1.756 | | | | | | |
| PBE | Zigzag | 3.298 | 14.89 | 3.971 | 11.538 | 0.716 | 17.044 | 3.642 | 2.220 | 2.259 | 3.545 |
| | Armchair | 4.625 | 2.42 | 5.261 | 4.419 | | | | | | |
| rPBE | Zigzag | 3.313 | 11.81 | 3.834 | 7.976 | 0.391 | 16.569 | 3.763 | 2.228 | 2.273 | 3.66 |
| | Armchair | 4.753 | 2.31 | 4.652 | 3.955 | | | | | | |
| PBEsol | Zigzag | 3.282 | 5.91 | 4.267 | 4.104 | 1.603 | 17.769 | 3.369 | 2.228 | 2.272 | 3.66 |
| | Armchair | 4.436 | 1.06 | 7.025 | 2.035 | | | | | | |
| PW91 | Zigzag | 3.304 | 7.22 | 3.965 | 4.585 | 0.502 | 17.039 | 3.668 | 2.234 | 2.262 | 3.545 |
| | Armchair | 4.625 | 1.62 | 5.223 | 2.448 | | | | | | |
| USPP_GGA | Zigzag | 3.301 | 0.51 | 3.951 | 0.298 | 2.251 | 17.084 | 3.677 | 2.220 | 2.260 | 3.548 |
| | Armchair | 4.626 | 0.15 | 5.249 | 0.192 | | | | | | |
| USPP_LDA | Zigzag | 3.264 | 5.81 | 4.346 | 3.541 | 0.209 | 18.123 | 3.338 | 2.196 | 2.224 | 3.329 |
| | Armchair | 4.371 | 1.59 | 8.828 | 2.274 | | | | | | |
| optB88 | Zigzag | 3.322 | 30.48 | 3.961 | 21.611 | 0.332 | 17.185 | 3.612 | 2.235 | 2.275 | 3.503 |
| | Armchair | 4.580 | 5.11 | 5.440 | 8.276 | | | | | | |
| optB86b | Zigzag | 3.304 | 15.93 | 4.077 | 10.717 | 0.179 | 17.476 | 3.458 | 2.263 | 2.259 | 3.438 |
| | Armchair | 4.506 | 3.26 | 6.063 | 3.314 | | | | | | |
| optPBE | Zigzag | 3.323 | 23.19 | 3.904 | 17.968 | 0.938 | 17.017 | 3.722 | 2.236 | 2.278 | 3.545 |
| | Armchair | 4.627 | 5.24 | 5.164 | 7.409 | | | | | | |
| vdW_DF2 | Zigzag | 3.378 | 14.56 | 3.771 | 10.328 | 0.074 | 16.443 | 4.156 | 2.263 | 2.321 | 3.685 |
| | Armchair | 4.782 | 2.41 | 4.589 | 4.104 | | | | | | |
| vdW_DF22 | Zigzag | 3.378 | 13.67 | 3.771 | 9.246 | 0.868 | 16.445 | 4.189 | 2.263 | 2.321 | 3.685 |
| | Armchair | 4.782 | 1.98 | 4.589 | 3.673 | | | | | | |

**Conflict of interest**

The authors declared that there is no conflict of interest.

**Acknowledgments**

G.Q. is supported by the National Natural Science Foundation of China (Grant No. 52006057), the Fundamental Research Funds for the Central Universities (Grant Nos. 531118010471 and 541109010001), and the Changsha Municipal Natural Science Foundation (Grant No. kq2014034). H.W. is supported by the National Natural Science Foundation of China (Grant No. 51906097). X.Z. is supported by the Fundamental Research Funds for the Central Universities (Grant No. 531118010490) and the National Natural Science Foundation of China (Grant No. 52006059). The numerical calculations in this paper have been done on the supercomputing system of the National Supercomputing Center in Changsha.